\documentclass[sigconf, nonacm]{acmart}
\usepackage{CJKutf8}

\AtBeginDocument{%
  \providecommand\BibTeX{{%
    \normalfont B\kern-0.5em{\scshape i\kern-0.25em b}\kern-0.8em\TeX}}}

\usepackage[linesnumbered,ruled,vlined]{algorithm2e}
\usepackage{multirow}
\newcommand{\tabincell}[2]{\begin{tabular}{@{}#1@{}}#2\end{tabular}}
\usepackage{array}
\newcommand{\PreserveBackslash}[1]{\let\temp=\\#1\let\\=\temp}
\newcolumntype{C}[1]{>{\PreserveBackslash\centering}p{#1}}
\newcolumntype{R}[1]{>{\PreserveBackslash\raggedleft}p{#1}}
\newcolumntype{L}[1]{>{\PreserveBackslash\raggedright}p{#1}}
\usepackage{booktabs}
\begin{document}
\begin{CJK}{UTF8}{gbsn}

\title{Retrieve Synonymous keywords for Frequent Queries in Sponsored Search in a Data Augmentation Way}

\author{Yijiang Lian}
\email{lianyijiang@baidu.com}
\affiliation{%
  \institution{Baidu}
}

\author{Zhenjun You}
\email{youzhenjun@baidu.com}
\affiliation{%
  \institution{Baidu}
}

\author{Fan Wu}
\email{wufan\_aais@pku.edu.cn}
\affiliation{%
  \institution{Peking University}
}

\author{Wenqiang Liu}
\email{liuwenqiang01@baidu.com}
\affiliation{%
  \institution{Baidu}
}

\author{Jing Jia}
\email{jiajing01@baidu.com}
\affiliation{%
  \institution{Baidu}
}




\begin{abstract}
In sponsored search, retrieving synonymous \emph{keywords} is of great importance for accurately targeted advertising. The semantic gap between queries and \emph{keywords} and the extremely high precision requirements (>= 95\%) are two major challenges to this task. To the best of our knowledge, the problem has not been openly discussed. In an industrial sponsored search system,
the retrieved \emph{keywords} for frequent queries are usually done ahead of time and stored in a lookup table. Considering these results as a seed dataset, we propose a data-augmentation-like framework to improve the synonymous retrieval performance for these frequent queries.
This framework comprises two steps: translation-based retrieval and discriminant-based filtering.
Firstly, we devise a Trie-based translation model to make a data increment. In this phase,
a Bag-of-Core-Words trick is conducted, which increased the data increment's volume 4.2 times while keeping the original precision.
Then we use a BERT-based discriminant model to filter out nonsynonymous pairs, which exceeds
the traditional feature-driven GBDT model with 11\% absolute AUC improvement.
This method has been successfully applied to Baidu's sponsored search system, which has yielded a significant improvement in revenue. In addition, a commercial Chinese dataset containing 500K synonymous pairs with a precision of 95\% is released to the public for paraphrase study\footnote{http://ai.baidu.com/broad/subordinate?dataset=paraphrasing}.

\end{abstract}


\keywords{Sponsored search, keyword matching, keyword retrieval, synonymous keywords retrieval, paraphrase generation, paraphrase identification}

\maketitle
\section{Introduction}
Sponsored search is one of the most major forms of online advertising and
also the main source of revenue for most search engine companies.
In sponsored search, there are three distinct roles involved: advertiser, user and search engine.
Each advertiser submits ads (abbreviated for advertisements)  and bids on a list of relevant \emph{keywords} for each ad.
To avoid ambiguity, \emph{keywords} in this paper is particularly used to denote queries purchased by advertisers.
When the search engine receives a query submitted by a user, it firstly retrieves a set of matched \emph{keywords}.
Then an auction is carried out to rank all corresponding ads,
taking both the quality and bid price of each ad into account.
Finally, the winning ads are presented on the search result page to the user.
\begin{figure*}[t]
  \centering
  \resizebox{1.0\totalheight}{!}{
      \includegraphics[width=1.7\textwidth]{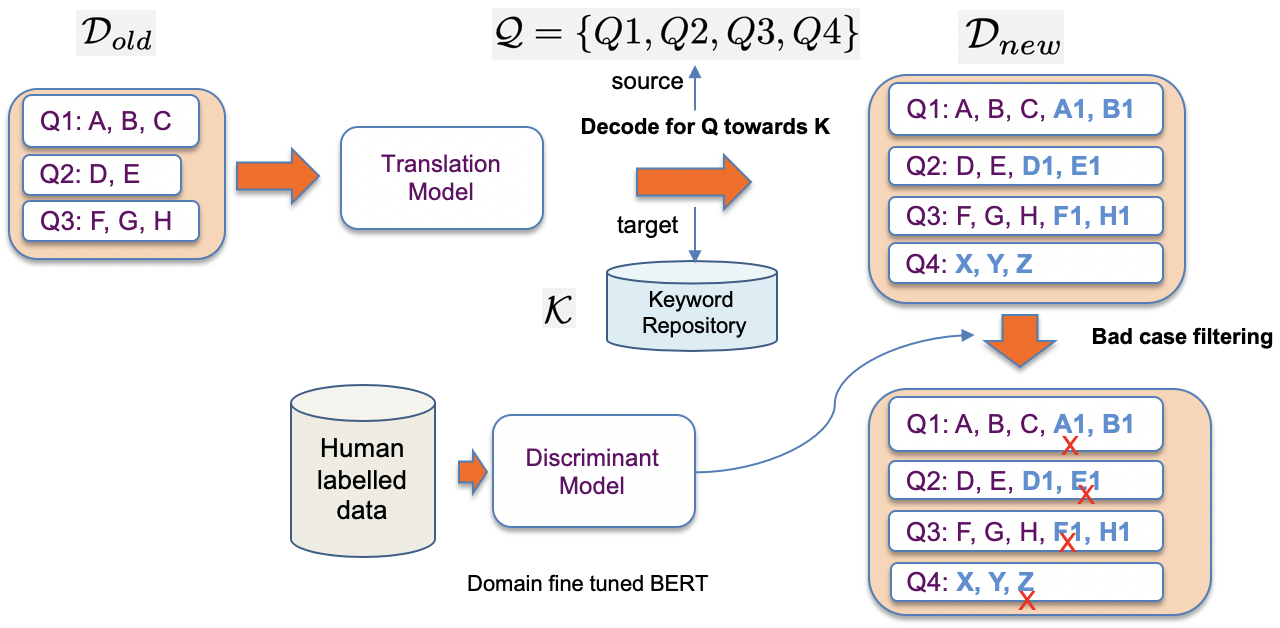}
       }
   \caption{Our framework contains two steps. Firstly, a Bag-of-Core-Words translation model is trained on $\mathcal{D}_{old}$, and is used to make constrained decoding for frequent query set $\mathcal{Q}$ towards the \emph{keyword} repository $\mathcal{K}$, which yields $\mathcal{D}_{new}$. Secondly, a BERT based synonym identification model is used to filter out nonsynonymous cases.}
  \label{fig:main_idea}
    \vspace{-3mm}
\end{figure*}

Major search engine companies provide a structured bidding language, with which the advertisers
can specify how would their purchased \emph{keywords} be matched
to the online queries. In general, 3 \emph{match types} are
supported:
\emph{exact, phrase, broad}.
In the early days, \emph{exact match} requires that query and \emph{keyword} are exactly the same.
Since queries are ever-changing, advertisers usually have to come up with a lot of
synonymous forms of their \emph{keywords} to capture more similar query flows. To ease their burden, modern
search engines relax the \emph{exact match}'s matching requirement to the synonymous level \footnote{https://support.google.com/google-ads/answer/2497825?hl=en}, which means
under \emph{exact match type}, the ad would be
eligible to appear when a user searches for the specific \emph{keyword} or its synonymous variants.
For example, the \emph{keyword} \emph{how much is iPhone 11} would not only be matched to the identical
query but also be matched to other queries like \emph{the price of iPhone 11}.
In \emph{phrase match type}, the matched queries should include the \emph{keyword} or the synonymous variants of the keyword.
\emph{Broad match type} further relaxes the matching requirements to the semantic relevance level.

The highly accurate matching makes \emph{exact match type} greatly welcomed by most customers, and nowadays it still occupies a great portion of the \emph{keywords} revenue for most search engine companies.
In this paper, we focus on the synonymous \emph{keywords} matching problem under the \emph{exact match type}.
To make it clear, for a given query and a \emph{keyword} repository (which is a snapshot of all the purchased \emph{keywords}), we want to retrieve as more synonymous \emph{keywords} as possible, while keeping a high precision. (On the one hand, retrieving more qualified synonymous \emph{keywords} can provide a more competitive advertisement queue for the downstream auction system; on the other hand, from the point of advertisers' fairness, each advertiser's synonymous \emph{keywords} should be retrieved.)

The vocabulary mismatch between queries and \emph{keywords} and the high precision required in the commercial product are two major challenges. Besides, since the volume of \emph{keywords} and queries might reach billions, how to effectively detect the synonymous relationships between these huge numbers of queries and \emph{keywords} is not an easy task.

Almost all of the existing published work about \emph{keyword} matching focused on nonsynonymous matching scenarios happened in \emph{phrase match} and \emph{broad match} \cite{azimi2015ads, malekian2008optimizing, lian2019end}.  As far as we know, few works have tried to tackle the synonymous matching problem in \emph{exact match} scenario.

It is well known that search queries are highly skewed and exhibit a power-law distribution
\cite{powerlaw2001, Petersen2016}. Approximately 20\% of frequent queries
occupy 80\% of the query volume.
In an industrial environment, the retrieved synonymous \emph{keywords} for frequent queries are stored in a Key-Value lookup table, which is computed and updated in an offline mode.
 When an ad hoc query arrives, corresponding synonymous \emph{keywords} can be retrieved immediately by looking it up on the table. This pre-retrieval offline architecture provides the industry with a good experimental environment, where lots of web data mining methods and state-of-the-art complex techniques can be tried.

\begin{figure}[!h]
  \centering
  \resizebox{1.0\totalheight}{!}{
      \includegraphics[width=1.5\textwidth]{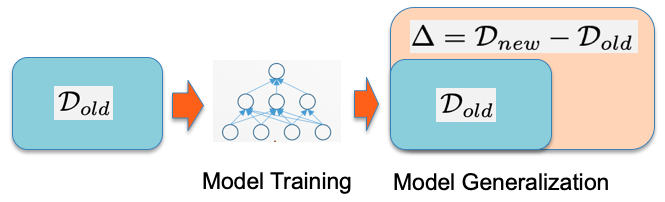}
       }
   \caption{A schematic diagram for the data augmentation framework.}
  \label{fig:data_aug}
    \vspace{-3mm}
\end{figure}

Considering the results stored on the lookup table as a dataset, previous iterations on this table provide us with a valuable data source. In this paper, we present a data-augmentation-like framework to improve the synonymous retrieval performance for the frequent queries, as is illustrated in Figure \ref{fig:data_aug}.
For brevity,
let's denote the frequent query set as $\mathcal{Q}$,
suppose we have already accumulated some high precision synonymous \emph{keywords}
for $\mathcal{Q}$ and denote the query-\emph{keyword} pair set as $\mathcal{D}_{old}$,
our motivation is to utilize the generalization capability of machine learning models to
expand $\mathcal{D}_{old}$ into $\mathcal{D}_{new}$, and $\Delta=\mathcal{D}_{new}-\mathcal{D}_{old}$ would be the new retrieved results.

Our framework contains two steps.
The first one is translation-based retrieval, where a translation model is trained to fit $\mathcal{D}_{old}$. Then
constrained decoding \cite{lian2019end} is conducted for $\mathcal{Q}$ towards the \emph{keywords} repository $\mathcal{K}$ to get $\mathcal{D}_{new}$.
To encourage the model to generate a larger $\Delta$, a synonym keeping Bag-of-Core-Words transformation is applied to the source and target side of $\mathcal{D}_{old}$.
The second step is bad case filtering. The translation model's
generalization ability could increase recall, but might also
introduce bad cases. To remove them, a strong synonym identification model based on
BERT \cite{bert} is introduced to score the sentence pairs in $\mathcal{D}_{new}$. Pairs with scores lower than a given threshold are filtered out.
As is shown in Figure \ref{fig:main_idea},
after translation-based retrieval, new \emph{keywords} A1, B1 are retrieved and appended after Q1's retrieval list [A, B, C]. Then a discriminant-based filtering is conducted, and  \emph{keyword} A1 is finally removed from Q1's expanded retrieval list.

Our main contributions are two folds:

Firstly, a practical data-augmentation-like framework is proposed to address the synonymous \emph{keywords} retrieval problem under \emph{exact match type}, which includes translation-based retrieval and discrimi--nant-based filtering. The Bag-of-Core-Words transformation trick increases the $\Delta$ substantially while keeping the original precision. And the domain fine-tuned BERT's performance far exceeds the feature-driven GBDT model's. To the best of our knowledge, this is the first time to address this important commercial problem. This method has been successfully applied to Baidu's sponsored search, which yields a significant improvement in revenue.

Secondly, a high-quality Chinese commercial synonym data set containing 500K pairs has been published along with this paper, which might be used in paraphrasing or other similar tasks. As far as we know, this is the first published large scale high-quality Chinese paraphrase data set reaching a precision of 95\%.

\section{Related Work}
The semantic gap between users and advertisers
is the most challenging problem in synonymous \emph{keywords} retrieval.
It could lead to the failure of the standard inverted index-based retrieval and has been widely discussed.
Some works introduced query-query transformation (query rewriting) \cite{malekian2008optimizing, zhang2007query, grbovic2015context, chen2019rpm} and \emph{keyword}-\emph{keyword} transformation \cite{azimi2015ads}. Direct query-\emph{keyword} transformation has also been studied \cite{Lee_2018, lian2019end}. \cite{lian2019end} proposed a Trie-constrained translation method to
make sure all generated sentences are valid commercial \emph{keywords}.
However, all of the above work focused on nonsynonymous matching scenarios.

Our framework consists of two main parts: a paraphrase generation model and a paraphrase identification model.
Paraphrase generation (PG), a task of rephrasing a given sentence into another with the same semantic meaning, has been used in various Natural Language Processing applications,
such as query rewriting \cite{zukerman2002lexical}, semantic parsing \cite{berant2014semantic}, and question answering \cite{rinaldi2003exploiting}.
Traditionally, it has been addressed using
rule-based approaches \cite{mckeown1983paraphrasing,Zhao_2009}.
Statistical machine translation has been used in \cite{wubben-etal-2010-paraphrase}.
Recent advances in deep learning have led to more powerful data-driven
 approaches to this problem.
\cite{sokolov2019neural} applied neural machine translation for paraphrase generation to
improve Alexa's \emph{ASK} user experience.
The semantic augmented transformer seq2seq model has also been studied \cite{wang2019task}.
The Variational AutoEncoder based generation model is also a good option \cite{gupta2018deep,park2019paraphrase}.

Paraphrase identification (PI) aims to determine whether two natural language sentences have identical meanings.
With the growing trend of PI, many English paraphrasing datasets have been made for this task, such as Quora Question pairs, and a lot of works have been developed based on them \cite{sharma2019natural, chandra2020experiments}.
Traditional methods mainly made use of handcrafted
features \cite{shen2016bidirectional, wan2006using, madnani2012re, das2009paraphrase}.
Recently, deep neural network (DNN) architectures have played a part in PI tasks.
According to whether the inner interaction between a pair of sentences is modeled, there are mainly two types of methods.
The first  is encoder-based.
RAE \cite{socher2011dynamic} is a pioneer that introduced recursive AutoEncoders  to PI.
Both ARC-I \cite{hu2014convolutional} and Hybrid Siamese CNN \cite{nicosia2017accurate} adopted Siamese architecture introduced in \cite{bordes2014semantic} but used different loss functions.
The other is interaction-based. ARC-II \cite{hu2014convolutional} and IIN \cite{gong2017natural} utilized the interaction space between two sentences while \cite{pang2016text} viewed the similarity matrix between words in two sentences as an image and utilized a CNN to capture rich matching patterns.
Bi-CNN-MI \cite{yin2015convolutional}, ABCNN \cite{yin2016abcnn}, BiMPM \cite{wang2017bilateral} and GSMNN \cite{fan2018globalization} focused on the effect of introducing multi-granular and multi-direction matching.
\cite{zhang2019paws} showed that BERT \cite{bert} pre-trained on a large corpus and then fine-tuned with an additional layer worked quite well on PI tasks.

\section{Method}
Our method can be formalized as 2 main steps: translation-based retrieval and discriminant-based filtering.
The seed paraphrasing dataset needed for the augmentation framework can be
any dataset of short paraphrasing text pairs.
In this paper, we won't go into much detail about paraphrase extraction techniques.
Our augmentation framework aims to retrieve more \emph{keywords} for each query while ensuring high precision.

\subsection{Translation-Based Retrieval}
Our translation model follows the common sequence to sequence learning
encoder-decoder framework \cite{Sutskever2014}.
And we implemented it with the Transformer \cite{vaswani2017attention},
considering its state-of-the-art performance in learning long-range dependencies and capturing the semantic structure of the sentences.


To increase the translation model's retrieval efficiency, a prefix tree for the \emph{keyword} repository is built ahead of time, and all the beam search decoding is constrained on this prefix tree \cite{lian2019end}.
At each step of the beam search,
the prefix tree will directly give the valid suffix tokens following the current hypothesis path,
then a greedy top-N selection is performed within the legal tokens.
This technique makes sure all the generated sequences are valid \emph{keywords}.

\begin{table}[!ht]
\caption{Some typical trivial synonymous translations for \emph{How much does double eyelid surgery cost}.}
\label{tab:trivial_translation}
\vspace{-3mm}
\centering
\begin{tabular}{|l|}
\toprule
\emph{How much does double eyelid surgery cost generally}\\
\emph{How much does double eyelid surgery cost in general?} \\
\emph{How much does double eyelid surgery cost probably}\\
\bottomrule
\end{tabular}
\end{table}


Paraphrases extracted from web data usually include some trivial patterns: reordering of words, insertion of function words and punctuation. A simple implementation of the translation model might generate too many trivial paraphrases, as is shown in Table \ref{tab:trivial_translation}.
Common stop words removing method is too coarse to meet our need for synonym keeping, especially in Chinese.
We carefully designed a synonym preserving data reduction method called Bag-of-Core-Words (abbreviated as \emph{BCW}) transformation to reduce each sentence into a compact form without losing semantic information.

For each tokenized sentence, the \emph{BCW} transformation consists of two sequential steps:
\begin{enumerate}
\item \textbf{Core-Words Transformation}. Here we consider some part of speech (abbreviated as POS) tags (like interjections, modal particles, etc.) as redundant, which means removing it generally does  not change the query's intention. Table \ref{tab:POS_removing} lists the typical redundant POS tags. Tokens with these tags would be removed. It is worth noting that some of the POS removing rules might not be  universally applicable to other languages. For example, modal particles and interjections are quite common in Mandarin Chinese, however, these words are unusual in English. And the remaining tokens are considered as core words.
For brevity, we refer to this step as \emph{CW} in the following sections.


\begin{table}[h]
\caption{Redundant part of speech tags in Chinese.}
\label{tab:POS_removing}
\vspace{-3mm}
\centering
\begin{tabular}{ll}
\toprule
\textbf{POS tag} & \textbf{Typical terms} \\
\midrule
Interjection &	哼(humph), 嗯 (em), 嘿 (hey), 嘘 (shh) \\
Auxiliary word &	等等 (and so on), 一般 (generally) \\
Punctuation &	comma, colon, question mark \\
Modal  particle &	啊 (ah), 哇 (wow), 呦 (yo), 耶 (yeah) \\\bottomrule
\end{tabular}
\end{table}

\item \textbf{Bag-of-Words Transformation}. In most cases word order does not affect the meaning of a sentence.
  In fact, we sample 600 commercial queries from the ad weblog and find that in 94\% of cases, the original query and its Bag-of-Words form
  have the same meaning. Based on this consideration, we sort the remaining core tokens in the sentence literally to remove the order's effect in the model training process. To make it more accurate, additional rules have been made to exempt the special cases. For example, \emph{Flights from New York to Beijing}
        and \emph{Flights from Beijing to New York} have the same Bag-of-words form, but their meanings are different.
        The same is true for \emph{Does hypertension cause hyperlipidemia} and \emph{Does hyperlipidemia cause hypertension}.
        So when sentences have two location tokens or two disease entity tokens or other token pairs with causality, temporal relation, etc.,
        the order of the paired tokens remains unchanged.
\end{enumerate}


The \emph{BCW} transformation is simple, fast and very effective.
Applying it to our Chinese training data effectively reduces the dataset size by nearly 20\% with a synonym precision of 98\%. Based on the \emph{BCW} transformation, we devised a data-augmentation-like method to retrieve synonymous \emph{keywords}, as is illustrated in
Algorithm \ref{2-step}.
\begin{algorithm}[!h]
\DontPrintSemicolon
\SetAlgoLined
\SetNoFillComment
            \caption{Retrieve synonymous \emph{keywords} in a data augmentation way.}
            \label{2-step}
            \KwIn{Synonymous query-\emph{keyword} pair dataset $\mathcal{D}_{old}$, \emph{keyword} repository $\mathcal{K}$,
            frequent queries set $\mathcal{Q}$, beam size $B$}
            \KwOut{Expanded query-\emph{keyword} pair dataset $\mathcal{D}_{new}$ }
            Apply the \emph{BCW} transformation to $\mathcal{D}_{old}$ to get $\mathcal{\widetilde{D}}_{old}$\;
            Train a neural machine translation model $M$ on $\mathcal{\widetilde{D}}_{old}$\;
            Apply the \emph{BCW} transformation to $\mathcal{K}$ to get $\mathcal{\widetilde{K}}$ and the corresponding relationship for elements in $\widetilde{\mathcal{K}}$ and $\mathcal{K}$ is stored in a lookup table $T$\;
         	Set the expanded dataset $\mathcal{D}_{new}$ to be a copy of $\mathcal{D}_{old}$ \;
            \For{each $q$ in  $\mathcal{Q}$}{
                Apply the \emph{BCW} transformation to $q$ to get $\widetilde{q}$\;
       		    Using $M$ to translate $\widetilde{q}$ towards $\mathcal{\widetilde{K}}$   with a beam size of $B$, which results retrieved \emph{keywords} set $\mathcal{R}_{\widetilde{q}}$ \;
		        Using $T$ to make inverse \emph{BCW} transformation on $\mathcal{R}_{\widetilde{q}}$, which results $\mathcal{R}_q$\;
                \For{each $k$ in  $\mathcal{R}_q$}{
                          Merge query-\emph{keyword} pair $<q,k>$ into $\mathcal{D}_{new}$\;
                }
        }
\end{algorithm}

\subsection{Discriminant-Based Filtering}
In business applications like sponsored search's matching product, high precision is essential.
The BERT-based classifier we use to further filter out bad cases has a similar structure with the classifier used in the sentence pair classification task illustrated in \cite{bert}. So we skip the exhaustive background description of the architecture and the training progress of the underlying model.


\section{Experiments}
\subsection{Dataset}
\noindent{\textbf{Dataset for translation model.}} The seed data $\mathcal{D}_{old}$
is extracted by calculating query-query similarity based on same URL
click-through information from the search engine's weblog \cite{zhao2010paraphrasing},
\emph{keyword}-\emph{keyword} similarity based on same advertiser purchase information
from the ad database and synonyms replacement. This data
is splitted into 3 parts: $\mathcal{D}_{train}^{PG}$ for training,
$\mathcal{D}_{dev}^{PG}$ for developing, and $\mathcal{D}_{test}^{PG}$ for testing.
The detailed statistics are shown in Table \ref{tab:translation_data_stat}.
The \emph{keyword} repository $\mathcal{K}$ contains 102,025,475 \emph{keywords}.
\begin{table}[!h]
    \vspace{-3mm}
\centering
    \caption{Statistics of datasets for the translation model.}
\label{tab:translation_data_stat}
\vspace{-3mm}
\begin{tabular}{cccc}
\toprule
    \textbf{Dataset} & \textbf{Query Number} & \textbf{Total Pairs} & \tabincell{c}{\textbf{Average Pairs}\\ (/query)} \\
\hline
\specialrule{0em}{1.25pt}{1.25pt}
    $\mathcal{D}_{train}^{PG}$ & 16,578,545  &110,687,827  &  6.67 \\
\specialrule{0em}{1.25pt}{1.25pt}
    $\mathcal{D}_{dev}^{PG}$ & 93,467 & 613,143 & 6.56  \\
\specialrule{0em}{1.25pt}{1.25pt}
    $\mathcal{D}_{test}^{PG}$ & 99,588 & 656,602 & 6.59  \\
\bottomrule
\end{tabular}
\vspace{-2.5mm}
\end{table}

\begin{table*}[!h]
\centering
    \caption{Results of different strategies.}
\label{tab:translation_recall}
\vspace{-3mm}
\begin{tabular}{ccccccc}
\toprule
    \textbf{Strategy} & \textbf{Beam Size} & \textbf{Diff ratio} & \textbf{BLEU-2} & \textbf{Dist-1/2}
    & \textbf{Precision} & \tabincell{c}{\textbf{Decoding Time}\\(ms/query)}  \\
\hline
    \emph{BASE-M} &30 &17.089\% & 0.446 & 0.0014/0.047& 83.5\% & 262.5   \\
    \emph{BASE-M} &120 & 69.753\%  & 0.356 & 0.0007/0.032 & 62.0\%  & 2988.9 \\
    \emph{CW-M} & 30 & 40.510\%  & 0.404 & 0.0015/0.058 & 83.0\% & 238.8 \\
    \emph{BCW-M} & 30 & 72.522\%  & 0.387 & 0.0018/0.061 & 82.5\% & 254.1  \\
\bottomrule
\end{tabular}
    \vspace{-2mm}
\end{table*}

\noindent{\textbf{Dataset for discriminant model.}}
To make a balanced domain dataset for human evaluation, three kinds of query-\emph{keyword} matching weblogs $\mathcal{D}_{exact}$, $\mathcal{D}_{phrase}$, $\mathcal{D}_{broad}$ are used, which correspond to the \emph{exact match}, \emph{phrase match}, and \emph{broad match} respectively. Among them, $\mathcal{D}_{exact}$ is probably synonymous and provides potential positive examples, while $\mathcal{D}_{phrase}$ and $\mathcal{D}_{broad}$ mainly contribute negative examples.
These three data sources are merged with the proportion of 2:1:1.
Then 170,000 data denoted as $\mathcal{D}_h^{PI}$ is sampled from it and sent to professionals for synonymous
binary human evaluation.
According to the human labels, 42.8\% of $\mathcal{D}_h^{PI}$ are positive samples and 57.2\% are negative.
$\mathcal{D}_h^{PI}$ is further split into three parts: 90\% of it is used as the domain specific data for BERT fine-tuning,
which is denoted as $\mathcal{D}_{train}^{PI}$; and 5\% of it is used for development,
denoted as $\mathcal{D}_{dev}^{PI}$ and the remaining 5\% denoted as $\mathcal{D}_{test}^{PI}$ is used for testing.

\subsection{Implementation Details}
For the translation model,
the word embeddings are randomly initialized. The vocabulary contains 100,000 most frequent tokens in the training data $\mathcal{D}_{train}^{PG}$.
The word embedding dimension and the number of hidden units are both set to 512.
For multi-layer and multi-head architecture of the Transformer, 4 encoder and decoder layers and 8 multi-attention heads are used.
And during the training, all layers are regularized with a dropout rate of 0.2.
And the model's cross-entropy loss is minimized with an initial learning rate of $5\times10^{-5}$
by Adam \cite{kingma2014adam} with a batch size of 128.

The paraphrase discriminant model is implemented
with BERT \cite{bert}, which takes a query-\emph{keyword} pair separated by a special token as input and predicts a synonymy label.
The model contains 12 layers, 12 self-attention heads, and the hidden dimension size is 768.
We initialize it with ERNIE \cite{ernie_baidu}, which learns Chinese lexical, syntactic and semantic information from a number of pretrained tasks, and fine-tuned it on $\mathcal{D}_{train}^{PI}$.
The fine-tuned loss is minimized with an initial learning rate of $1 \times 10^{-6}$
by Adam \cite{kingma2014adam} with a batch size of 64. We evaluate our model after each epoch and stop training when the validation loss on $\mathcal{D}_{dev}^{PI}$ does not decrease after 3 epochs.

All of the experiments are run on a machine equipped
with a 12-core Intel(R) Xeon(R) E5-2620 v3 clocked at 2.40GHz, a
RAM of 256G and 8 Tesla K40m GPUs.

\subsection{Results of Translation Model}

For the translation model, we compared the retrieval performances of 3 different strategies.
For all of the strategies, decoding is conducted in a Trie-constrained mode. For the sake of convenience, these strategies are abbreviated as follows:
\begin{itemize}
\item \emph{BASE-M}: This is our base strategy, where the translation model is trained on
  the original training data  $\mathcal{D}_{train}^{PG}$. And the prefix tree is built on $\mathcal{K}$.
\item \emph{CW-M}: In this strategy, the model is trained on the \emph{CW}-transformed data. And the prefix tree is built on \emph{CW}-trans--formed $\mathcal{K}$. The final result is made by joining the generated hypotheses with $\mathcal{K}$ based on the \emph{CW} transformation.
\item \emph{BCW-M}: In this strategy, the model is trained on the \emph{BCW}-transformed data. And the prefix tree is built on \emph{BCW}-trans--formed $\mathcal{K}$. The final result is made by joining the generated hypotheses with $\mathcal{K}$ based on the \emph{BCW} transformation.
\end{itemize}

Each of the trained models is utilized to decode
towards those queries in $\mathcal{D}_{test}^{PG}$ to get a result data set $\mathcal{D}_{1}^{PG}$.
Under the data augmentation framework, we expect the translation model could make a large
data increment $\Delta$
while keeping a high precision. There are three major concerns: the size of $\Delta$, the precision of $\Delta$, and the decoding time for generating $\Delta$. The following indicators are considered for evaluation:
\begin{itemize}
    \item \textbf{Diff ratio} is defined as $\frac{|\mathcal{D}_{1}^{PG}-\mathcal{D}_{test}^{PG}|}{|\mathcal{D}_{test}^{PG}|}$, which is an indicator of the generalization ability.
    \item \textbf{BLEU-n} is an indirect indicator of the generation quality. Since most sentences in our scenario are short texts, we only consider BLEU-2.
    \item \textbf{Dist-n} measures the number of distinct N-grams within the set of generated data, which indicates the diversity among the generated paraphrases. Following previous studies \cite{park2019paraphrase}, we use Dist-1,2.
    \item \textbf{Precision} indicates the proportion of synonymous pairs in generated data. Concretely,
      for each strategy, 400 query-\emph{keyword} pairs are sampled from the generated results $\mathcal{D}^{PG}_1$
      for binary human evaluation. We denote this dataset as $\mathcal{D}^{PG}_{1h}$ for later reference.
\end{itemize}

Rows 1, 3 and 4 in Table \ref{tab:translation_recall} show the retrieval performances
for these three different strategies with a beam size of 30.
We can see that: \emph{BASE-M} could already make a certain amount of Diff ratio, which proves the feasibility of the data-augmentation-like framework. \emph{CW-M} and \emph{BCW-M} further enlarge the Diff ratio to 2.4 times and 4.2 times, compared with the base method. Meanwhile, the precision of \emph{CW-M} and \emph{BCW-M} are almost the same as that of \emph{BASE-M}. The Dist indicator shows that the results' diversity has been improved by \emph{CW-M} and \emph{BCW-M}.

For further analysis, we evaluate the decoding results of \emph{BASE-M} with a beam size of 120.
As is shown in Table \ref{tab:translation_recall}, although the diff ratio increases up to 4 times, which is nearly equal to \emph{BCW-M} with a beam size of 30, the precision and diversity drop significantly.
What's more, \emph{BASE-M} has to spend more than 10 times of time to make it.

To conclude, \emph{BCW-M} greatly enlarges the Diff ratio  while maintaining a high level precision. It is fast and almost little extra time is consumed.

\subsection{Results of Discriminant Model}
Our baseline model is a GBDT (Gradient Boosting Decision Tree) model \cite{ke2017lightgbm} trained on a collection of human designed text similarity features.
The features are listed as follows:
\begin{enumerate}
\item Token level matching degree: max matching length, the proportion of matching and missing tokens, BM25, BLEU1 and BLEU2.
\item Named entity similarity: whether the named entities in query and \emph{keyword} are matched.
\item Simple Approximate Bigram Kernel \cite{ozatecs2016sentence} based on dependency parsing tree.
\item Document class similarity: whether query and \emph{keyword} belong to the same document class.
\item Semantic similarity: DSSM \cite{huang2013learning} trained with query-\emph{keyword} click-noclick pairs shown in the weblog, and Word2vec based cosine similarity \cite{mikolov2013efficient}.
\item Translation likelihood: the \emph{BASE-M} translation score which is calculated by $P(\emph{keyword}~|~\rm{query})$.
\end{enumerate}

GBDT is also trained on $\mathcal{D}^{PI}_{train}$, and validated on $\mathcal{D}^{PI}_{dev}$. And the model hyperparameters are optimized by grid search.
For evaluation, we use two metrics: the area under an ROC curve (AUC) and recall under 95\% precision.
Table \ref{tab:human_judge_diff} shows the models' performances  on $\mathcal{D}_{test}^{PI}$.
To our surprise, BERT greatly exceeds GBDT's performance. For the AUC indicator, BERT outperforms GBDT
by 11.4 percentage points. Under the precision of 95\%, BERT's recall exceeds GBDT by 45 percentage points.

\begin{table}[h]
\centering
    \caption{Model performances of \emph{BERT} VS \emph{GBDT}  on $\mathcal{D}_{test}^{PI}$. \textbf{Recall} indicates the recall ratio under the precision of 95\%.}
\label{tab:human_judge_diff}
\vspace{-3mm}
\begin{tabular}{cccc}
\toprule
    \textbf{Strategy} & \textbf{AUC} & \textbf{Recall} \\
\hline
\emph{GBDT} & 84.4\% &  21.8\% \\
\emph{BERT} & 95.8\% &  66.8\% \\
\bottomrule
\end{tabular}
\vspace{-2mm}
\end{table}

\begin{table}[h]
    \vspace{-2mm}
\centering
    \caption{Model performances on $\mathcal{D}_{1h}^{PG}$ generated by \emph{BCW-M}. \textbf{Recall} indicates the recall ratio under the precision of 95\%. }
\label{tab:recall_on_d_new}
\vspace{-4.5mm}
\begin{tabular}{ccc}
\toprule
    \textbf{Strategy} & \textbf{AUC} & \textbf{Recall}   \\
\hline
\emph{GBDT} & 75.0\% & 25.2\% \\
\emph{BERT} & 94.7\% & 91.3\% \\
\bottomrule
\end{tabular}
\end{table}

\begin{table*}[tp]
\small
\centering
    \caption{Top-5 beam search results of \emph{BASE-M} VS \emph{BCW-M}.}
   \label{tab:case_bcwt}
    \vspace{-3.5mm}
    \begin{tabular}{L{1.5cm}cc|L{1.5cm}cc}
    \toprule
       \textbf{Query} &  \textbf{\emph{BASE-M}}  & \textbf{\emph{BCW-M}}
        & \textbf{Query} & \textbf{\emph{BASE-M}} & \textbf{\emph{BCW-M}}\\
    \midrule
        \specialrule{0em}{0.5pt}{0.5pt}
        \multirow{5}*{\tabincell{l}{金 的 市场 价格\\ \tabincell{l}{Market price\\ of gold} }}
        & \tabincell{c}{金 市场 价格\\Gold price in the market}
        & \tabincell{c}{价格| 金| 市场\\ gold| market|price}
        & \multirow{5}*{\tabincell{l}{化妆 的 培训班\\ \tabincell{l}{Training school\\ of makeup}}}
        & \tabincell{c}{化妆 培训 学校 哪家 好\\Which makeup training school is good?}
        &  \tabincell{c}{化妆 |培训| 学校\\makeup | school| training } \\
        ~ & \tabincell{c}{价格| 金| 市场\\gold| market|price}
        & \tabincell{c}{多少| 黄金| 钱\\how much|gold}
        & ~ & \tabincell{c}{化妆 培训 学校\\Makeup training school}
        &  \tabincell{c}{班 |化妆| 学习\\course |learning| makeup}\\
        ~ & \tabincell{c}{金 价格\\Gold  price}
        & \tabincell{c}{价格| 金\\price|gold}
        & ~ &\tabincell{c}{化妆 的 培训 学校\\Training school of makeup}
        & \tabincell{c}{化妆 |机构| 培训\\institution | makeup| training} \\
        ~ & \tabincell{c}{金 的 价格\\Price of gold}
        & \tabincell{c}{查询| 价格| 金\\gold |price| query}
        & ~ &\tabincell{c}{化妆 培训 的 学校\\School of makeup training}
        & \tabincell{c}{彩妆 |培训| 学校\\cosmetic| school| training} \\
        ~ & \tabincell{c}{市场 金 价格\\Market gold price}
        & \tabincell{c}{黄金| 价格| 最新\\current |gold | price}
        & ~ & \tabincell{c}{培训 化妆 的 学校\\School of training makeup}
        & \tabincell{c}{化妆 |学| 学校\\ learning| makeup| school}\\
    \bottomrule
\end{tabular}
\end{table*}
We also
considered the models' performances on $\mathcal{D}^{PG}_{1}$ generated by \emph{BCW-M}, since our motivation is to remove bad cases in the translations.
Here we use the previously human evaluated dataset $\mathcal{D}^{PG}_{1h}$ for testing.
As is shown in Table \ref{tab:recall_on_d_new}, BERT achieves a recall of 91.3\% under the precision of 95\%, which also
far exceeds the GBDT's performance.
(The difference of recall ratios in Table \ref{tab:recall_on_d_new} and Table \ref{tab:human_judge_diff} comes from the
difference of data distribution between $\mathcal{D}^{PG}_{1h}$ and $\mathcal{D}_{test}^{PI}$. In $\mathcal{D}^{PG}_{1h}$, positive examples account for 82.5\% ,
while in $\mathcal{D}_{test}^{PI}$, positive examples account for 42.8\%.)

BERT's dramatic improvement might come from the following points: a)
BERT has a super large parameter space in contrast to the simple semantic similarity features like Word2vec and DSSM. The huge capacity
and abundant pretraining make BERT being able to learn a lot of external semantic similarity knowledge, which is especially useful to alleviate
the vocabulary mismatch in paraphrase identification.
b) The paraphrase relationship between two sentences is usually judged by the alignment of their tokens.
To some extent, the multi-head attention mechanism in the Transformer might make soft alignments in multiple semantic spaces.


\subsection{Case Study and Discussion}
\subsubsection{Bag-of-Core-Words Transformation}
Table \ref{tab:case_bcwt} shows some typical cases for the \emph{BCW-M} strategy. Due to space constraints, we
only present the top 5 decoding results. The second column in Table \ref{tab:case_bcwt} shows the
final results of \emph{BASE-M}, and the last column shows the raw generated results of \emph{BCW-M}
without joining the \emph{keyword} repository, which is presented as Bag-of-Words form and
the symbol `|' is used as the token separator. We can see that with \emph{BCW-M},  the number of
redundant translations decreases a lot. For example, before \emph{BCW-M} is applied, most of the
generated hypotheses for query (金的市场价格, market price of gold) are quite trivial: 金市场价格 just
removes the auxiliary token 的, and 市场金价格 simply further reorders these tokens. When \emph{BCW-M} is applied,
nontrivial bag-of-words paraphrases like (多少|黄金|钱, Gold| how much) and (查询|价格|黄金, Gold |price| query)
emerged into the top 5 hypotheses list. Similarly, for query (化妆的培训学校, Training school of makeup),
four nearly identical translations (化妆培训学校,化妆的培训学校，化妆培训的学校，培训化妆的学校) are generated under
the base strategy. When \emph{BCW-M} is applied, the results are not dull anymore.

\subsubsection{Good Cases and Bad Cases}

\begin{table*}[!ht]
   \centering
    \caption{Some typical synonymous \emph{keywords} generated by the translation model.}
   \label{tab:good_bad_cases}
    \vspace{-3.5mm}
    \begin{tabular}{llc}
 \toprule
       \textbf{Query}  & \textbf{Generated \emph{Keywords}} & \textbf{Label} \\
 \midrule
男士减肥机构   & 男性瘦身机构 & Good\\
Men's Weight Loss Agency & Agency for men slimming & Good \\
厨房油污怎样清除       & 厨房油渍如何去除 & Good \\
How to clean off oil in the kitchen &  How to remove kitchen greasy dirt & Good \\
宝宝身上胎记是怎么回事  &婴儿身上胎记是什么原因 & Good \\
What causes a birthmark on a baby & What is the reason of a birthmark & Good \\
怎么在跑步机上跑步 & 怎么跑步减肥 & Bad \\
How to run on treadmill & How to lose weight through running  & Bad \\
 \bottomrule
   \end{tabular}
 \end{table*}
Table \ref{tab:good_bad_cases} shows some sampled results generated by our translation model.
We can see that in most cases our model greatly captures the synonym relationship between query and \emph{keyword}.
The vocabulary mismatch problem has been alleviated to some extent. For example, the synonym relationships between (减肥, weight loss) and (瘦身, slimming), (油污, oil) and (油渍, greasy dirt), etc. have been captured. Some complex paraphrases like (宝宝身上胎记是怎么回事, 婴儿身上胎记是什么原因) (What causes a birthmark on a baby, what is the reason of a birthmark) have been generated, which could not be simply accomplished by synonymous phrase replacement.

In the meanwhile, bad cases might also be generated. Sometimes, important intentions in the query  might be discarded. For example, in the case of (怎么在跑步机上跑步, 怎么跑步减肥) (How to run on a treadmill, How to lose weight through running), the intention of `treadmill' is lost.

\subsubsection{Why Translation?}
People might argue that since we have trained a discriminant model, why not use this model to do direct retrieval? In other words, can we make retrieval by simply scoring all the query-\emph{keyword} pairs in the Cartesian product set?
The reason is the scalability problem in the real industry environment mentioned in the introduction. The \emph{keyword} volume might reach a magnitude of billions. And the frequent query set size is usually set to reach 10 million to have a great impact on the revenue. Computing so many pairs directly with a BERT model is impossible even in an offline mode, for it costs about $3.5\times10^7$ days on 8 Tesla K40m GPUs to score all pairs.
Translation based method gives us a practical way to find more confident query-\emph{keyword} pairs which are more likely to be synonym candidates, thus greatly narrowing down the possible candidates' size.


\subsection{The published Dataset}
Our work has a close relationship with paraphrase generation and paraphrase identification.
Most of the existing large scale paraphrasing datasets (MSCOCO, SNLI, etc.) are in English and high-quality Chinese dataset is extremely scarce in this domain. The most related dataset is a 24K sized dataset LCQMC \cite{liu2018lcqmc}. However, it focuses on general intent matching rather than paraphrasing. To promote the Chinese paraphrasing research, we decide to publish a large scale high-quality dataset containing 500K commercial synonymous short text pairs along with this paper.
This dataset is produced in the following steps:
A translation model trained with the \emph{BASE-M} strategy is used to make unconstrained decoding for real-world frequent queries, where the prefix tree constraint is discarded. Then our discriminant model is used to filter out bad cases. Finally, some heuristic rules are devised to filter out trivial translations. Manual sampling evaluation shows that this dataset has a precision of 95\%.

\subsection{Online Experiments}
A real online A/B test experiment is deployed on Baidu's commercial advertising system, where two fractions of  search
flow are sent to the experimental group and the control group independently.
$\mathcal{D}_{old}$ is used as the Key-Value table in the control group and  $\mathcal{D}_{old}+ \Delta$ in the experimental group.
We use 10,000,000 frequent queries, and the translation model is trained with \emph{BCW-M} strategy.

For each group, $\#\{\rm{searches}\}$ is used to denote the total number of queries it received, $\#\{\rm{clicks}\}$ to denote the corresponding number of clicks, and $\rm{revenue}$ to denote the search company revenue.
The following metrics are calculated independently for each group to evaluate the performance of our method.

$\bullet$ SHOW denotes the total number of ads shown to users.

$\bullet$ $\rm{CTR}=\frac{\#\{\rm{clicks}\}}{\#\{\rm{searches}\}}$, which
denotes the average clicks received by the search  engine.

$\bullet$ $\rm{ACP}=\frac{\sum\rm{price}}{\#\{\rm{clicks}\}}=\frac{\rm{revenue}}{\#\{\rm{clicks}\}}$, which denotes
the average click price paid by the advertisers.

$\bullet$ $\rm{CPM}=\frac{\rm{revenue}}{\#\{\rm
  {searches}\}}\times 1000=\mathrm{CTR}\times \mathrm{ACP}\times 1000$, which
denotes the average revenue received by the search engine for 1000
searches.

$\bullet$ Quality refers to the query-\emph{keyword} synonym relationship. For each side of
this A/B experiment, 600 query-\emph{keyword} cases under the \emph{exact match type} (excluding literally identical ones)
are sampled from the system's ad weblog and are sent for binary human evaluation.  And the quality score
is calculated as the proportion of the synonymous pairs.

\begin{table}[h]
\vspace{-3mm}
  \caption{Online A/B Test performance of our method.}
  \label{tab:online_evalu}
\vspace{-3mm}
  \centering
\centering
\begin{tabular}{ccccc}
\toprule
\textbf{SHOW} & \textbf{CTR} & \textbf{ACP} & \textbf{CPM}  & \textbf{Quality} \\
\midrule
 +0.8\% & +1.02\%& +0.62\%& +1.64\%  & +1.2\% \\
\bottomrule
\end{tabular}
\vspace{-2.5mm}
\end{table}


  Table \ref{tab:online_evalu} shows the relative improvements in these indicators. We can see that the incremental dataset $\Delta$ increases the CPM by 1.64\%,
  which is a significant improvement for our revenue.
  We give our intuitive explanation from the demand-supply perspective.
  The sponsored search system aims to match query flows to the advertisers' demands. Detecting more synonym relationships between queries and  \emph{keywords} helps to get more ads into the downstream auction phase.
  The ACP's 0.62\% growth shows the competition in the ad queue has been intensified.
  Finally, the number of shown ads increased by 0.8\%, and the clicks increased by 1.02\%.
  The growth in clicks and prices combined to bring about the final CPM growth.
  In the meanwhile, the human evaluation demonstrates that
  the ads' relevance quality has not been deteriorated.



\section{Conclusions}
In this paper, we present a simple but effective data-augmentation-like framework
to address the synonymous \emph{keyword} retrieval problem for frequent queries in sponsored search.
Based on a seed paraphrasing dataset, 
a sequence to sequence translation model is trained and used
to decode out more synonymous pairs to \emph{expand} the data.
To ensure high precision for commercial usage, a domain fine-tuned BERT is used to filter out bad cases.
During the translation phase, we introduce a novel scheme to make the decoding more effective:
Bag-of-Core-words transformation, which enlarges the diff 4.2 times while almost keeping the original precision.
During the discrimination phase, BERT outperforms the traditional feature-driven GBDT model by 11 percentage points.
To the best of our knowledge, this is the first published work to address the synonymous \emph{keywords} retrieval problem. Our framework has been successfully applied in Baidu's sponsored search.
As a byproduct, 500K  high quality commercial Chinese synonymous pairs have been published along with this paper. To the best of our knowledge, this is also the first published large scale Chinese paraphrasing dataset with a precision of 95\%.
\bibliographystyle{ACM-Reference-Format}
\bibliography{paraphrase.bib}
\end{CJK}
\end{document}